\documentclass[a4paper,fleqn,usenatbib]{mnras}

\usepackage{newtxtext,newtxmath}

\usepackage[T1]{fontenc}
\usepackage{ae,aecompl}

\usepackage{graphicx}	
\usepackage{amsmath}	
\usepackage{amssymb}	
\usepackage{epsfig}

\title[HERUS: The Far-IR/Submm SEDs of Local ULIRGs]
{HERUS: The Far-IR/Submm Spectral Energy Distributions of Local ULIRGs \& Photometric Atlas\thanks{Based in part on observations with {\em Herschel}, an ESA space observatory with science instruments provided by European-led Principal Investigator consortia and with important participation from NASA}}
\author[D.L.~Clements et al.]
{ \parbox{\textwidth}{\raggedright D.L.~Clements$^{1}$\thanks{E-mail: \texttt{d.clements@imperial.ac.uk}},
C. Pearson$^{2,3,4}$,
D.~Farrah$^{5}$,
J.~Greenslade$^1$,
Jeronimo Bernard-Salas$^3$,
E. Gonz\'alez-Alfonso$^6$,
J. Afonso$^{7,8}$,
A. Efstathiou$^9$,
D. Rigopoulou$^{2,4}$,
V. Lebouteiller$^{10,11}$,
P. D. Hurley$^{12}$,
H. Spoon$^{13}$
\vspace{0.4cm}\\
\parbox{\textwidth}{\raggedright{\small
$^{1}$Astrophysics Group, Imperial College London, Blackett Laboratory, Prince Consort Road, London SW7 2AZ, UK\\
$^2$RAL Space, CCLRC, Rutherford Appleton Laboratory, Chilton, Didcot, Oxfordshire OX11 0QX, UK\\
$^3$School of Physical Sciences, The Open University, Milton Keynes, MK7 6AA,
UK\\
$^4$Oxford Astrophysics, Denys Wilkinson Building, University of Oxford, Keble
Rd, Oxford OX1 3RH, UK\\
$^{5}$Department of Physics, Virginia Tech, Blacksburg, VA 24061, USA\\
$^6$Universidad de Alcal\'a, Departamento de F\'{\i}sica y Matem\'aticas, Campus Universitario, E-28871 Alcal\'a de Henares, Madrid, Spain\\
$^7$Instituto de Astrof\'{i}sica e Ci\^{e}ncias do Espa\c co, Universidade de Lisboa, OAL, Tapada da Ajuda, PT1349-018 Lisboa, Portugal\\
$^8$Departamento de F\'{i}sica, Faculdade de Ci\^{e}ncias, Universidade de Lisboa, Edif\'{i}cio C8, Campo Grande, PT1749-016 Lisbon, Portugal\\
$^9$School of Sciences, European University Cyprus, Diogenes Street, Engomi, 1516, Nicosia, Cyprus\\
$^{10}$Universit{\'e} Paris Diderot, AIM, Sorbonne Paris Cit{\'e}, CEA, CNRS, F-91191 Gif-sur-Yvette, France\\
$^{11}$IRFU, CEA, Universit{\'e} Paris-Saclay, F-91191 Gif-sur-Yvette, France\\
$^{12}$Astronomy Centre, Department of Physics and Astronomy, University of Sussex, Falmer, Brighton BN1 9QH, UK\\
$^{13}$Space Sciences Building, Cornell University, Ithaca, NY 14853, USA}}}
\date{}
\pagerange{\pageref{firstpage}--\pageref{lastpage}}
\pubyear{}
}
\begin{document}

\maketitle

\label{firstpage}

\begin{abstract}
We present the {\em Herschel}-SPIRE photometric atlas for a complete flux limited sample of 43 local Ultraluminous Infrared Galaxies (ULIRGs), selected at 60$\mu$m by {\em IRAS}, as part of the HERschel ULIRG Survey (HERUS). Photometry observations were obtained using the SPIRE instrument at 250, 350 and 500$\mu$m. We describe these observations, present the results, and combine the new observations with data from IRAS to examine the far-IR spectral energy distributions (SEDs) of these sources. We fit the observed SEDs of HERUS objects with a simple parameterised modified black body model where temperature and emissivity $\beta$ are free parameters. We compare the fitted values to those of non-ULIRG local galaxies, and find, in agreement with earlier results, that HERUS ULIRGs have warmer dust (median temperature $T=37.9\pm4.7$ K compared to $21.3 \pm 3.4$ K) but a similar $\beta$ distribution (median $\beta = 1.7$ compared to 1.8) to the {\em Herschel} reference sample (HRS, Cortese et al., 2014) galaxies. Dust masses are found to be in the range of $10^{7.5}$ to $10^{9}$ M$_{\odot}$, significantly higher than that of Herschel Reference Sample (HRS) sources.  We compare our results for local ULIRGs with higher redshift samples selected at 250 and 850$\mu$m. These latter sources generally have cooler dust and/or redder 100-to-250 $\mu$m colours than our 60$\mu$m-selected ULIRGs. We show that this difference may in part be the result of the sources being selected at different wavelengths rather than being a simple indication of rapid evolution in the properties of the population. 


\end{abstract}


\section{Introduction}

Ultraluminous Infrared Galaxies (ULIRGs) are a rare class of objects in the local universe which have a far-IR luminosity (8 to 1000$\mu$m) $> 10^{12} L_{\odot}$. They were initially discovered by the {\em IRAS} satellite (Wright et al., 1984;  Soifer et al., 1984) and were then extensively studied through ground-based followup (eg. Sanders et al., 1988a) and by subsequent space missions (see Sanders \& Mirabel, 1996 and references therein). Almost all local ULIRGs turn out to be the result of a merger between two gas rich galaxies (eg. Clements et al., 1996, Rigopoulou et al., 1999, Farrah et al., 2001) and are powered by a major starburst or AGN activity triggered by this interaction. They have also been suggested as the progenitors of quasars (eg. Sanders et al., 1988b), the central engines of which would be powered by the same gas reservoir that powers their starbursts. While rare in the local universe, these objects evolve strongly with redshift even at low redshifts, from $z\sim0$ to $z\sim0.3$ (Kim et al., 1998). The advent of submm imagers and far-IR space missions subsequent to {\em IRAS} have shown that higher redshift equivalents of ULIRGs are at least 2 orders of magnitude more common at $z\sim$ 2 than they are in the local universe (eg. Hughes et al., 1998; Smail et al., 1997; Dole et al., 2001; Gruppioni et al., 2013). Such sources have been implicated in the formation of massive cluster galaxies at high redshift (Farrah et al., 2006), the development of the Cosmic Infrared Background (Dole et al., 2005), and the AGN-triggered quenching of star formation in massive galaxies (Hopkins, 2012).

Understanding the physical processes behind the numerous ULIRG-class objects at high redshift is not easy - their distance means they are faint despite their luminosity. This means that we must instead use the local ULIRG population to provide guidance in understanding the more distant objects. Once we understand the local sources we can then apply this knowledge to the more distant population under the assumption that they are similar enough that local insights can set a baseline for comparison, and detection of any evolution with redshift. ULIRGs are extremely dusty objects, containing typical dust masses of $10^{8} M_{\odot}$ or greater. Their most luminous regions can thus be heavily obscured (see eg. Rangwala et al., 2011). A detailed view of their most obscured regions must thus be obtained at long wavelengths, in the far-IR or submm, since shorter wavelength radiation cannot penetrate the obscuring material. While there have been a number of studies of individual ULIRGs at these wavelengths in the past (eg. Fischer et al., 2010; Rangwala et al., 2011, Fischer et al., 2014), it is only through the statistical study of large, complete samples that we will obtain clear insights into the breadth of ULIRG properties and behaviours. This is the role of the {\em Herschel} ULIRG Survey (HERUS), the largest coherent spectroscopic and photometric survey of local ULIRGs to date.

Previous HERUS papers have largely dealt with the spectroscopic observations of the sample (Pearson et al., 2016; Farrah et al., 2013; Spoon et al., 2013), or results on individual objects, such as IRAS 08572+3915 (Efstathiou et al., 2014). The present paper presents the photometric observations of the HERUS ULIRGs obtained using the SPIRE instrument (Griffin et al., 2010) on the {\em Herschel} Space Observatory (Pilbratt et al., 2010), and results derived from them. These include the results of fitting simple parametric SEDs to the {\em IRAS} and HERUS data for these objects and comparing these results to those from similar analysis of other data sets. The simple $(T, \beta)$ SED model (definition given in Section 5) we use is widely applied in this field (see eg. Cortese et al., 2014, Clements et al., 2010, Dunne et al., 2000). This model is clearly a major simplification of the emission of actual dust in a real galaxy, since there are many other factors, such as dust at multiple temperatures, optical depth effects (eg. Rangwala et al., 2011), the evolution of dust properties with time and metallicity, and more that can apply. This can affect the interpretation of such simplified SED fits. Cortese et al. (2014), for example, find that the choice of a fixed $\beta$ value can strongly affect the derived dust temperature and the correlation of that temperature with other galaxy properties. Fortunately our current data set provides enough information for $\beta$ to be a free parameter that is determined during the fitting process, avoiding at least some of these problems. Optical depth effects will tend to flatten the SED around its peak, so that a simple $(T, \beta)$ model is no longer a good fit, while additional dust components, whether warmer or cooler will also modify the SED and the interpretation of the results. Since ULIRGs are complex and very dusty objects, with strong evidence that at least some are optically thick even at far IR wavelengths (Rangwala et al., 2011), some of these issues are likely to arise in our data. The physical properties of dust will evolve over time and may well be affected by the strong radiation fields that arise during the massive starbursts and AGN activity that powers a ULIRG. This may be reflected in the results of our SED fits and systematic differences between what we find for ULIRGs and what other authors find for more quiescent galaxies.

The paper is structured as follows: in the next section we describe the HERUS sample, a complete sample of ULIRGs in the local universe at $z<0.3$ and with 60$\mu$m flux $>$1.8Jy. In section 3 we describe the SPIRE observations and in the subsequent section we describe the data reduction and photometric results obtained. In section 5 we use the SPIRE photometry combined with {\em IRAS} fluxes for these sources to fit a simple parametric dust SED model to the data. From this we derive ULIRG dust temperatures, emissivities and masses, and compare these to other far-IR sources in the literature. In section 6 we look at the residuals to our fits to examine whether there are any systematic issues with our simple model across the whole sample and to see what improvements might be made to this model, compare the far-IR colours of the HERUS ULIRGs to other far-IR samples, and examine the selection effects that might affect the different samples of ULIRG-like objects selected by {\em IRAS} in the local universe, and by {\em Herschel} and/or submm imagers at higher redshift. We summarise our conclusions in section 7. The SPIRE images of the HERUS ULIRGs and plots of their SED fits are presented in the appendices. We assume a Hubble constant of $H_0 = 70$ kms$^{-1}$ Mpc$^{-1}$ and density parameters of $\Omega_M = 0.3$ and $\Omega_L = 0.7$.

\section{The HERUS Sample}

The {\em Herschel} ULIRG Survey (HERUS, PI D. Farrah, programme ID OT1\_dfarrah\_1) was, at 250 hours, among the largest open time extragalactic programmes carried out by the {\em Herschel Space Observatory}. It is an unbiased survey of local ULIRGs comprising a sample of 43 objects selected to be at $z<0.3$, with 60$\mu$m fluxes $>$1.8Jy and originally identified in the IRAS PSC-z survey (Saunders et al., 2000).

The entire sample was observed by {\em Herschel} (Pillbratt et al., 2010) using the SPIRE instrument (Griffin et al., 2010) in both photometer mode, at 250, 350 and 500$\mu$m, and in Fourier Transform Spectrometer (FTS) mode. The results of the latter observations are reported in Pearson et al. (2016). The sources were also observed at shorter far-IR wavelengths using the PACS instrument (Poglitsch et al., 2010). These observations were split between the HERUS survey (Spoon et al., 2013; Farrah et al., 2013) and the SHINING survey (Fischer et al., 2010; Sturm et al., 2011; Veilleux et al., 2013; Gonzalez-Alfonso et al., 2015, 2017; Janssen et al., 2016; ). Other observations for this sample are also available at a range of resolutions and wavelengths. For more details on the SPIRE spectroscopic observations see Pearson et al. (2016).

\section{SPIRE Observations}

The HERUS SPIRE photometry observations were carried out between 16th August 2011 ({\it Herschel} Operational Day (OD) 825) and 16th May 2012 (OD 1022). The observations are summarised in Table \ref{tab:observations} where the source name, redshift and infrared luminosity (L$_{IR}$= L$_{8-1000\umu m}$) are tabulated along with the {\it Herschel}-SPIRE observation I.D.s ({\it obsid}) for the corresponding photometric observations and their OD.

The SPIRE photometer observations were carried out in Small Map mode ({\it SpirePhotoSmallScan}, POF10, Dowell et al., 2010) with  fixed, 3 repetition (445s), cross-linked 1$\times$1 scan legs covering a field of 4$\arcmin$ radius. Images are taken simultaneously in the SPIRE  250$\umu$m (PSW), 350$\umu$m (PMW), and 500$\umu$m (PLW) bands.  Note that 3C273  only has photometric data and the photometric data for IRAS 06035-7102 was extracted from the Open Time Key Programme: KPOT$\_$mmeixner observations of the Large Magellanic Cloud, Level 2.5  {\it Herschel} data product. 

\begin{table*}
\caption{Summary of HERUS observations of local ULIRGs. The operational day (OD) and observation identification (obsID) are tabulated for  the observations made with the SPIRE photometer.  Indicative far-infrared luminosities ( L$_{IR}$= L$_{8-1000\umu m}$) calculated following Sanders \& Mirabel (1996) are also included for reference.}
\begin{tabular}{@{}lllll} \hline
Target Name & Redshift & log(L$_{IR}$/L$_{\sun}$) & OD& ObsID\\
\hline
IRAS00397-1312 & 0.262 & 12.97 & 949 & 1342234696 	\\
Mrk1014        & 0.163 & 12.61 & 976 & 1342237540 	\\
3C273          & 0.158 & 12.72 & 948 & 1342234882\\
IRAS03521+0028 & 0.152 & 12.56 & 1022 & 1342239850 	\\
IRAS07598+6508 & 0.148 & 12.46 & 862 & 1342229642 	\\
IRAS10378+1109 & 0.136 & 12.38 & 948 & 1342234867 \\
IRAS03158+4227 & 0.134 & 12.55 & 825 & 1342226656 \\
IRAS16090-0139 & 0.134 & 12.54 & 862 & 1342229565 \\
IRAS20100-4156 & 0.13 & 12.63 & 880 & 1342230817 \\
IRAS23253-5415 & 0.13 & 12.26 & 949 & 1342234737 \\
IRAS00188-0856 & 0.128 & 12.47 & 949 & 1342234693 \\
IRAS12071-0444 & 0.128 & 12.31 & 948 & 1342234858 \\
IRAS13451+1232 & 0.122 & 12.32 & 948 & 1342234792$^{*}$\\
IRAS01003-2238 & 0.118 & 12.25 & 949 & 1342234707 \\
IRAS11095-0238 & 0.107 & 12.28 & 948 & 1342234863 \\
IRAS20087-0308 & 0.106 & 12.41 & 880 & 1342230838 \\
IRAS23230-6926 & 0.106 & 12.32 & 880 & 1342230806 \\
IRAS08311-2459 & 0.1 & 12.46 & 880 & 1342230796 \\
IRAS15462-0450 & 0.099 & 12.24 & 989 & 1342238307 \\
IRAS06206-6315 & 0.092 & 12.23 & 825 & 1342226638 \\
IRAS20414-1651 & 0.087 & 12.24 & 892 & 1342231345 \\
IRAS19297-0406 & 0.086 & 12.38 & 880 & 1342230837 \\
IRAS14348-1447 & 0.083 & 12.33 & 989 & 1342238301 \\
IRAS06035-7102 & 0.079 & 12.14 & 353 & 1342195728$^{*}$ \\
IRAS22491-1808 & 0.078 & 12.18 & 949 & 1342234671 \\
IRAS14378-3651 & 0.067 & 12.14 & 989 & 1342238295 \\
IRAS23365+3604 & 0.064 & 12.17 & 948 & 1342234919 \\
IRAS19254-7245 & 0.062 & 12.06 & 515 & 1342206210$^{*}$ \\
IRAS09022-3615 & 0.06 & 12.24 & 880 & 1342230799 \\
IRAS08572+3915 & 0.058 & 12.16 & 880 & 1342230749 \\
IRAS15250+3609 & 0.055 & 12.03 & 948 & 1342234775 \\
Mrk463         & 0.05 & 11.77 & 963 & 1342236151 \\
IRAS23128-5919 & 0.045 & 12.06 & 544 & 1342209299$^{*}$ \\
IRAS05189-2524 & 0.043 & 12.12 & 467 & 1342203632$^{*}$ \\
IRAS10565+2448 & 0.043 & 12.08 & 948 & 1342234869 \\
IRAS17208-0014 & 0.043 & 12.38 & 467 & 1342203587$^{*}$ \\
IRAS20551-4250 & 0.043 & 12.01 & 880 & 1342230815 \\
Mrk231 & 0.042 & 12.49 & 209 & 1342201218$^{*}$ \\
UGC5101 & 0.039 & 12.01 & 495 & 1342204962$^{*}$ \\
Mrk273 & 0.038 & 12.13 & 438 & 1342201217$^{*}$ \\
IRAS13120-5453 & 0.031 & 12.22 & 829 & 1342226970 \\
NGC6240 & 0.024 & 11.93 & 467 & 1342203586$^{*}$ \\
Arp220 & 0.018 & 12.14 & 229 & 1342188687$^{*}$ \\
\hline
& \multicolumn{3}{c}{* denotes {\it Herschel} data obtained outside the HERUS project.}
\end{tabular}
\label{tab:observations}
\end{table*}

\section{Data Reduction and Results}

All SPIRE photometry observations (listed in Table ~\ref{tab:observations}) were  processed through the standard Small Map User Pipeline with HIPE 11.2825, using SPIRE Calibration Tree 11.0 with default values for all pipeline tasks. Target positions in the map were found by the HIPE SUSSEXtractor task (Savage \& Oliver, 2007), assuming a Full Width Half Maximum (FWHM) of 18.15, 25.2, 36.9 arcseconds  for the PSW, PMW, PLW bands respectively. These positions were then input into the SPIRE Timeline Fitter task within the HIPE environment (Pearson et al., 2014) that fits a Gaussian function to the baseline-subtracted SPIRE timelines. The background is measured within an annulus of between 300 and 350 arcsec and then an elliptical Gaussian function is fit to both the central 22, 32, 40 arcsec (for the PSW, PMW, PLW bands respectively) and the background annulus. The results for the photometry of all the HERUS galaxies is shown in Table ~\ref{tab:photometry} while the images in the three {\em Herschel} bands are included in Appendix A. Absolute flux calibration accuracy for SPIRE is estimated to be 4\% while relative flux errors are 1.5\% or better (Bendo et al., 2013).

A small number of sources appear to have close companions in the SPIRE images. These include IRAS 01003-2238 (80 arcsecond separation), IRAS14378-3651 (30 arcsecond separation), IRAS15462-0450 (35 arcsecond separation) and IRAS23253-5415 (27 arcsecond separation). In most cases the companions are detected in 2MASS with comparable magnitudes and usually somewhat bluer colours than the ULIRG itself. The conclusion of Duc et al. (1997) that the main component in IRAS15462-0450, which they classify as a loose interacting pair, is responsible for all the IRAS emission is perhaps challenged by the HERUS SPIRE images, which show that the component about 35 arcseconds to the S might contribute about 20\% of the flux. In contrast to the others, the companion $\sim$27 arcsec to the NE of IRAS23253-5415 is not detected by 2MASS but it is detected by WISE. The WISE images suggest that this companion and a weaker extension to the E may in fact all be part of the same system. A detailed analysis of the companions and a search for any weak extended far-IR emission in these systems, which are largely unresolved by SPIRE, is beyond the scope of this paper. 

\begin{table*}
\caption{Photometry of the HERUS galaxies in the SPIRE bands PSW (250$\umu$m), PMW (350$\umu$m), PLW (500$\umu$m). These fluxes originally appeared in Pearson et al. (2016)}
\centering
\begin{tabular}{@{}ccccccc} \hline
\hline
Target &   \multicolumn{2}{l}{F250} &  \multicolumn{2}{l}{F350}  &  \multicolumn{2}{l}{F500} \\
          &  Flux Density   & Error & Flux Density & Error & Flux Density & Error	\\
     &Jy&Jy&Jy&Jy&Jy&Jy\\
\hline
IRAS00397-1312  & 0.389 & 0.004 & 0.130 & 0.004 & 0.040 & 0.005	\\
Mrk1014         & 0.460 & 0.004 & 0.175 & 0.004 & 0.063 & 0.005	\\
3C273           & 0.437 & 0.004 & 0.633 & 0.004 & 0.994 & 0.005	\\
IRAS03521+0028  & 0.684 & 0.004 & 0.270 & 0.004 & 0.094 & 0.004	\\
IRAS07598+6508  & 0.500 & 0.004 & 0.197 & 0.004 & 0.058 & 0.005	\\
IRAS10378+1109  & 0.480 & 0.004 & 0.183 & 0.004 & 0.050 & 0.005	\\
IRAS03158+4227  & 0.973 & 0.004 & 0.377 & 0.004 & 0.137 & 0.005	\\
IRAS16090-0139  & 1.067 & 0.004 & 0.404 & 0.004 & 0.116 & 0.005	\\
IRAS20100-4156  & 1.001 & 0.004 & 0.349 & 0.004 & 0.102 & 0.005	\\
IRAS23253-5415  & 1.044 & 0.005 & 0.437 & 0.004 & 0.165 & 0.005	\\
IRAS00188-0856  & 0.877 & 0.004 & 0.345 & 0.004 & 0.111 & 0.005	\\
IRAS12071-0444  & 0.471 & 0.004 & 0.163 & 0.004 & 0.044 & 0.005	\\
IRAS13451+1232  & 0.503 & 0.005 & 0.256 & 0.004 & 0.197 & 0.006	\\
IRAS01003-2238  & 0.222 & 0.004 & 0.070 & 0.004 & 0.026 & 0.006	\\
IRAS23230-6926  & 0.617 & 0.004 & 0.204 & 0.004 & 0.064 & 0.005	\\
IRAS11095-0238  & 0.380 & 0.004 & 0.119 & 0.004 & 0.036 & 0.005	\\
IRAS20087-0308  & 1.804 & 0.006 & 0.687 & 0.004 & 0.210 & 0.005	\\
IRAS15462-0450  & 0.492 & 0.004 & 0.162 & 0.004 & 0.050 & 0.008	\\
IRAS08311-2459  & 1.246 & 0.005 & 0.464 & 0.004 & 0.148 & 0.005	\\
IRAS06206-6315  & 1.248 & 0.005 & 0.477 & 0.004 & 0.158 & 0.005	\\
IRAS20414-1651  & 1.315 & 0.005 & 0.519 & 0.004 & 0.168 & 0.005	\\
IRAS19297-0406  & 2.039 & 0.006 & 0.752 & 0.004 & 0.244 & 0.005	\\
IRAS14348-1447  & 1.842 & 0.006 & 0.666 & 0.005 & 0.197 & 0.006	\\
IRAS06035-7102  & 1.226 & 0.022 & 0.397 & 0.001 & 0.130 & 0.008	\\
IRAS22491-1808  & 0.862 & 0.004 & 0.305 & 0.004 & 0.097 & 0.005	\\
IRAS14378-3651  & 1.330 & 0.005 & 0.478 & 0.005 & 0.135 & 0.006	\\
IRAS23365+3604  & 1.849 & 0.006 & 0.669 & 0.004 & 0.210 & 0.005	\\
IRAS19254-7245  & 1.545 & 0.005 & 0.587 & 0.004 & 0.185 & 0.005	\\
IRAS09022-3615  & 2.449 & 0.007 & 0.823 & 0.004 & 0.252 & 0.005	\\
IRAS08572+3915  & 0.504 & 0.004 & 0.164 & 0.004 & 0.060 & 0.004	\\
IRAS15250+3609  & 0.966 & 0.004 & 0.368 & 0.004 & 0.136 & 0.005	\\
Mrk463          & 0.344 & 0.004 & 0.134 & 0.004 & 0.052 & 0.005	\\
IRAS23128-5919  & 1.565 & 0.008 & 0.556 & 0.006 & 0.176 & 0.007	\\
IRAS10565+2448  & 3.619 & 0.011 & 1.319 & 0.004 & 0.407 & 0.005	\\
IRAS20551-4250  & 1.629 & 0.005 & 0.556 & 0.004 & 0.170 & 0.005	\\
IRAS05189-2524  & 1.963 & 0.011 & 0.717 & 0.007 & 0.211 & 0.009	\\
IRAS17208-0014  & 7.918 & 0.037 & 2.953 & 0.010 & 0.954 & 0.009	\\
Mrk231          & 5.618 & 0.019 & 2.011 & 0.008 & 0.615 & 0.008	\\
UGC5101         & 6.071 & 0.039 & 2.327 & 0.018 & 0.746 & 0.009	\\
Mrk273          & 4.190 & 0.011 & 1.493 & 0.006 & 0.471 & 0.006	\\
IRAS13120-5453 & 12.097 & 0.036 & 4.441 & 0.010 & 1.355 & 0.006	\\
NGC6240         & 5.166 & 0.029 & 2.031 & 0.009 & 0.744 & 0.008	\\
Arp220          & 30.414 & 0.132 & 12.064 & 0.036 & 4.145 & 0.015	\\
\hline
\end{tabular}
\label{tab:photometry}
\end{table*}

\section{Dust Properties of the HERUS ULIRGs}

The far-IR properties of galaxies can be studied by fits of their far-IR to submm SEDs to models of various types. These can range from simple parameterised models of modified black bodies (eg. Clements et al., 2010; Cortese et al., 2014; Dunne et al., 2000; Vlahakis et al., 2005; Clements et al., 1993; Bendo et al., 2003 and references therein) to multicomponent dust models (eg. Clements et al., 2010; Klass et al., 2001; Dunne \& Eales 2001), to full-scale radiative transfer models that include dust of different compositions at a wide range of temperatures (eg. Farrah et al., 2003; Ciesla et al., 2014, Efstathiou \& Siebenmorgan 2009). Needless to say, the more complex the model, the more photometric points over a wider range of wavelengths are needed to properly constrain the fit. For the present photometric atlas, which includes just the combination of IRAS and SPIRE photometry from 60 to 500$\mu$m that is available for the complete dataset, we restrict ourselves to the simplest of the parametric models available, a modified black body function where the emission as a function of frequency, $\nu$ is given by:
\begin{equation}
F_{\nu}[\nu, T, \beta] \propto \nu^{\beta} B_{\nu}(\nu, T) 
\end{equation}
where $F_{\nu}$ is the flux density in Jy, $\beta$ is an emissivity parameter that typically has a value between 1 and 2, $T$ is a parameter specifying the dust temperature, and $B_{\nu}(\nu,T)$ is the standard black body function. We do not claim that the dust SED in the HERUS ULIRGs, or in fact any galaxy, is explained by dust of a single type at a single temperature, we simply use this particular commonly used SED parameterisation as a way of comparing the far-IR properties of the HERUS ULIRGs to those of other galaxy samples that have widely been fit using a similar parameterisation. Future papers (Farrah et al., in prep, Clements et al., in prep, Efstathiou et al., in prep) will explore other types of SED fits using a wider range of heterogenous data, including the off-line continuum fluxes from our PACS observations, as well as a range of more sophisticated parametric and physical models.

Once we have fitted the SEDs of our HERUS sources to this model we compare our results to those of a number of different studies of both ULIRGs and other galaxy samples in the literature. In Table {\ref{table:samples} we list these samples, their selection criteria, the basic parameters of the observations and statistics of their results as well as similar details for the current work.

\begin{table*}
\begin{tabular}{cccccc} \hline
Source&Selection&Complete?&Passbands Used ($\mu$m)&$\bar{T}$ (K)&$\bar{\beta}$\\ \hline
Current Work&ULIRGs, $z<0.3$, F$_{60}>1.8$Jy&Yes&60,100,250,350,500&37.9&1.8\\
Cortese et al., 2014&15$< D<25$Mpc, $K<12.5$ LTG, $K<8.7$ ETG&Yes&100, 160, 250, 350, 500&21.3&1.7\\
Clements et al., 2010&ULIRGs, F$_{60}>1$ Jy, 850$\mu$m detected ($z<0.19$)&No&60, 100, 850&42&1.6\\
Dunne et al., 2000&$v > 1900$ km/s F$_{60}>5.4$ Jy -10$<\delta<50$&Yes&60, 100, 850&35.6&1.3 \\
Smith et al., 2013&$z<0.5$ HATLAS DR1 region, F$_{250}>20.4$ mJy& Yes&100, 160, 250, 350, 500&23.5&1.82\\
Magdis et al., 2014&$0.22<z<0.9^{*}$, F$_{250} >150$mJy, HerMES region&Yes&70, 160, 250, 350, 500&36&1.5 fixed\\
Chapman et al., 2005&F$_{850}$>4.5 mJy, SCUBA surveys$^+$, 1.4 GHz detected&No&850, 1.4GHz&36&1.5 fixed\\
Yang et al., 2007&F$_60 > 0.2$Jy, $z$ known ($0.1<z<1$), $K$ known&No&60, 100, 350, 1.4GHz&42.8&1.5 fixed\\
Huang et al., 2014&$z>4$, 250 \& 350 detected&No&250, 350, 850, 1100&64&2.0 fixed\\ \hline

\end{tabular}
\caption{Comparison samples used in this study. Basic selection criteria are given but some more complex aspects, such as low galactic foreground, omitted for clarity. See the relevant papers for full details. The complete column indicates the whether the sample can be considered statistically complete. Fixed in the $\beta$ column indicates that this parameter was set to this value and not part of the fit. LTG = late type galaxies, ETG = early type galaxies. The HATLAS DR1 region amounts to 161 sq. deg. in three equatorial regions (Eales et al., 2010). $^*$two higher z lensed sources also included. The HerMES region amounts to 90 sq. deg. in a number of fields (Oliver et al., 2012). See Chapman et al. (2005) for details of the regions surveyed. Redshift range for these sources is 0.08 to 3.4. Fixed in the $\beta$ column indicates that this parameter was set to this value and not part of the fit. }
\label{table:samples}
\end{table*}

Two of our sources, 3C273 and IRAS 13451+1232 (aka 4C12.5) are dominated by non-thermal emission in the far-IR (Clements et al., 2010). Analysis of the properties of any dust emission in these objects thus requires a thorough knowledge of the non-thermal synchrotron spectrum in these sources which is beyond the scope of this paper. We thus exclude these objects from further consideration here.

\subsection{Fitting Method}

We adopt a Bayesian fitting method for the HERUS sources similar to that used for nearby galaxies observed by {\rm Planck} (Planck Collaboration, 2011). We start with a model of the data where the measured flux $d_{\nu}$ at frequency $\nu$ is of the form:
\begin{equation}
d_{\nu} = A F_{\nu}\left[ (1+z)\nu; T, \beta\right] + n_{\nu}
\end{equation}
where $A$ gives the overall amplitude of the SED and $(1+z)$ converts the rest frame frequency to the observed frequency of a given channel for an object at a redshift of $z$. The noise contribution at a given frequency $\nu$ is given by $n_{\nu}$ which we model as a Gaussian with variance $\sigma_{\nu}^2$ and which is determined by the {\em Herschel} or {\em IRAS} observations. The parameters of our model are thus $A, \beta$ and $T$. We fit these to the data for each individual HERUS ULIRG using the \verb!emcee! Markov Chain Monte Carlo package\footnote{ http://dan.iel.fm/emcee/current/} (Foreman-Mackey et al., 2013, for the general approach see also Lewis \& Bridle 2002, Jaynes 2003) to probe the parameter space, equivalent to an exploration of the $\chi^2$ surface with a nonlinear parameterization. We apply the following priors for our parameter values: for temperature ($T$) we adopt a uniform prior between 3 and 100K; for emissivity ($\beta$) we adopt a uniform prior between 0 and 3; for A, because the wide range of distances for our sources implies a range of amplitudes covering several orders of magnitude, we adopt a uniform prior on $\log(A)$ between -3 and 10. The probability is zero outside the range of these uniform prior distributions.

We start a series of MCMC chains at positions in the three parameter search space chosen at random from a uniform distribution with $T=30 - 50K, \beta = 1.2 - 1.8$ and $\log(A) = -1$ to $3$ to match our expectation for the likely range of these parameters. A hundred of these starting points were chosen and the MCMC engine was then used to generate 600 further steps along each of these chains for a total of 60000 samples. The first 100 steps in each of these chains is regarded as `burn in time' and the remaining 50000 samples are used to determine the best fit parameters and their posterior distributions. Fig \ref{fig:mcmc} and Fig \ref{fig:sed} show the sampling chain, the resulting two dimensional and marginalized likelihood functions, and the resulting SED fit for IRAS 00397-1312 to demonstrate this fitting process. The results of these fits, giving the median in the marginalized posterior distributions as the fitted value and the 32nd and 68th percentiles of the distribution as the 1$\sigma$ errors, are given in Table \ref{table:fits}.

One potential issue with our fitting method is that several of our main comparison samples use data from a different range of filters: we use IRAS+SPIRE at 60, 100, 250, 350 and 500 $\mu$m while the wholly {\em Herschel} derived samples use PACS+SPIRE at 100, 160, 250, 350, 500$\mu$m. We tested whether these differences might introduce biases into our results by taking a range of model SEDs of known temperature, extracting model fluxes at the relevant wavelengths, adding noise appropriate to the different sets of observations, and then seeing whether there were any significant deviations from a one-to-one fit. The only significant issue we find is a bias towards higher dust temperatures for fits using PACS+SPIRE data for input dust temperatures $>$50K, where the output temperature is high compared to that input. There were no substantial biases for the IRAS+SPIRE models comparable to the HERUS data. Since there are no PACS+SPIRE sources with $T>50K$ in our comparison samples we conclude that no significant biases are introduced to our analysis as a result of the slightly different wavelength bands being used.

\begin{figure*}
\epsfig{file=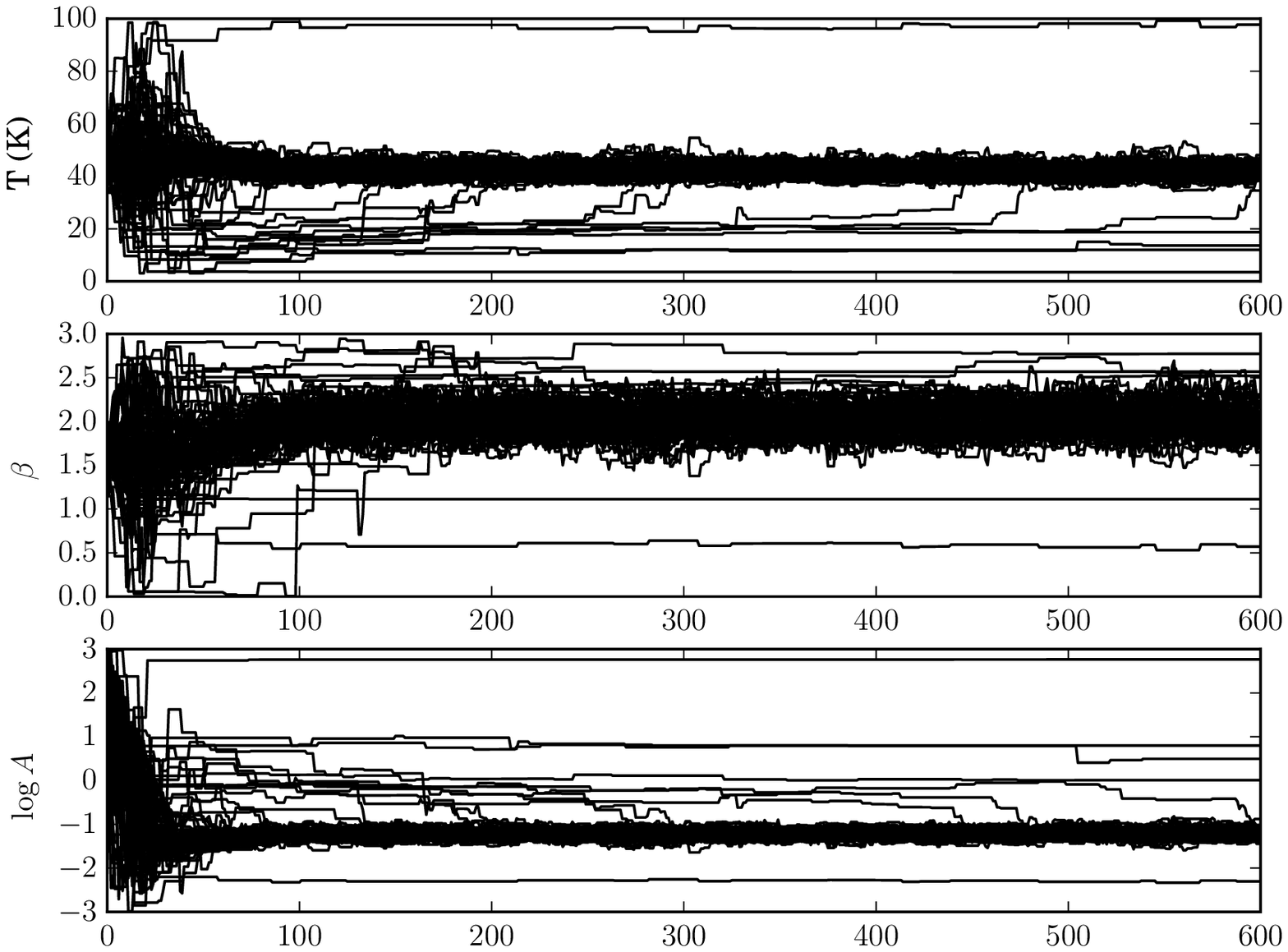,angle=0, width=10cm}
\epsfig{file=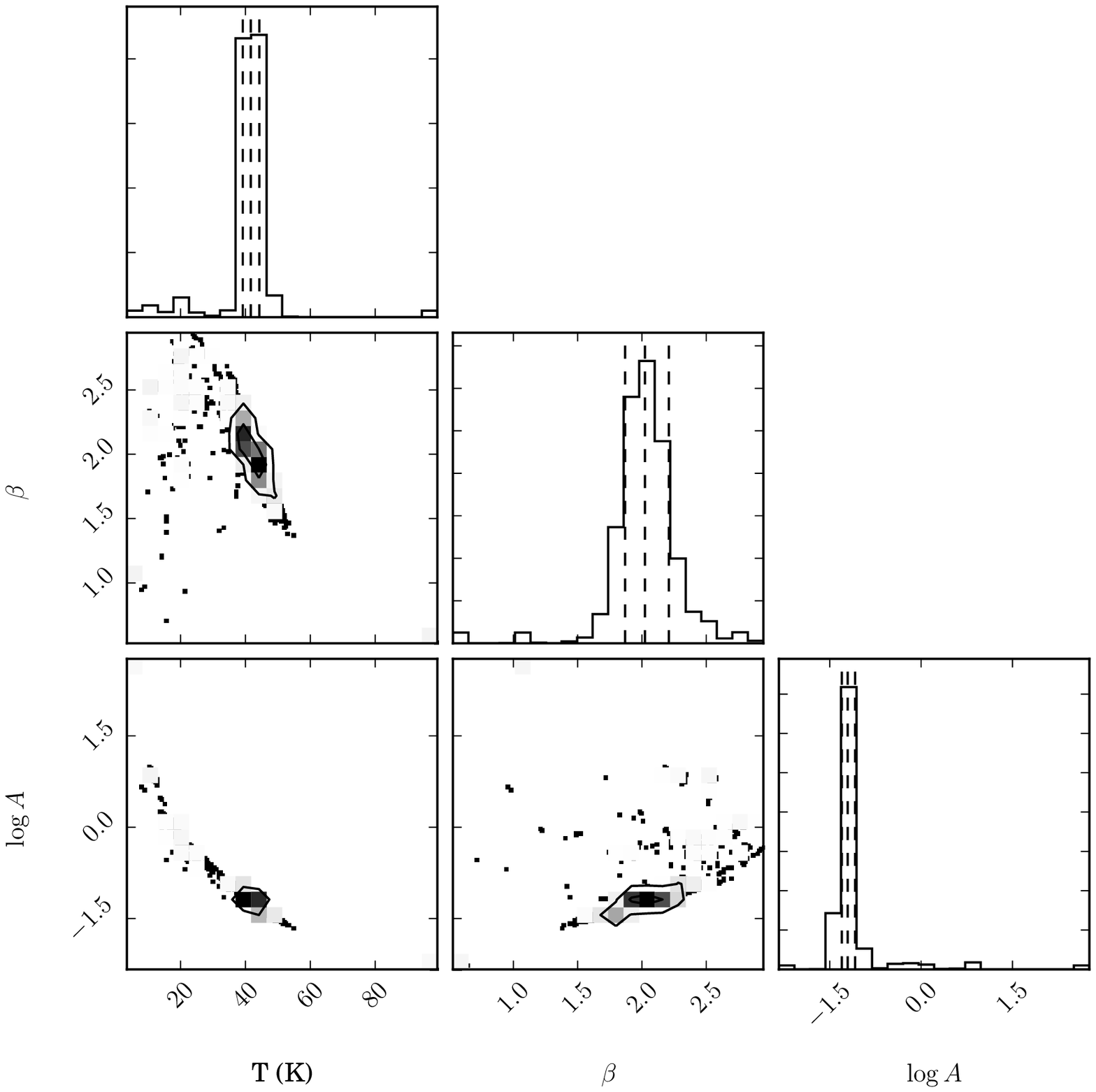,angle=0, width=10cm}
\caption{Demonstration of our MCMC fitting method as applied to the ULIRG IRAS00397-1312. {\bf Top:} The complete set of 100 MCMC chains including the first 100 steps of burn in period which are discarded. The convergence of nearly all the chains to very similar values demonstrates the reliability of the fitting process. {\bf Bottom:} The two dimensional and marginalized posterior likelihood functions showing that the SED parameters are very well fitted by this method. }
\label{fig:mcmc}
\end{figure*}

\begin{figure}
\epsfig{file=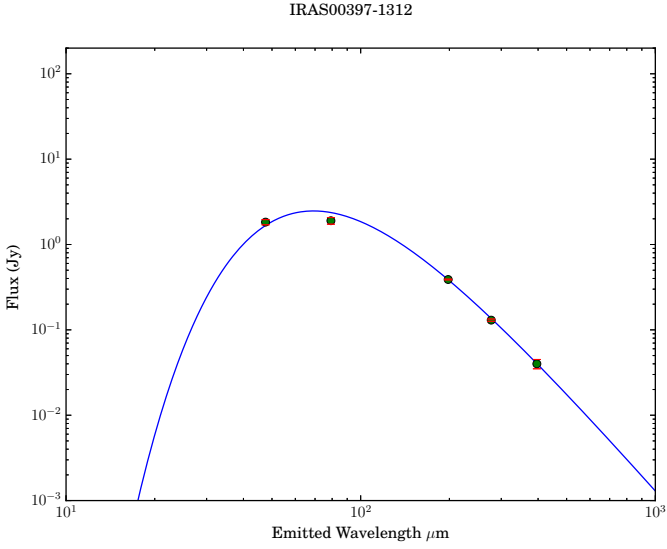,angle=0, width=10cm}
\caption{The final fitted SED for IRAS00397-1312 using the median set of marginalized parameters derived from the MCMC fitting process. Fit is shown as the blue line, data points as green dots. Error bars on the fluxes are also shown (in red) but are generally smaller than the green dots of the data points.}
\label{fig:sed}
\end{figure}

\begin{table*}
\begin{tabular}{|l|r|l|l|l|l|l|l|l|l|l|l|}
\hline
  \multicolumn{1}{|c|}{Name} &
  \multicolumn{1}{c|}{z} &
  \multicolumn{1}{c|}{Temp (K)} &
  \multicolumn{1}{c|}{T+} &
  \multicolumn{1}{c|}{T-} &
  \multicolumn{1}{c|}{Beta} &
  \multicolumn{1}{c|}{B+} &
  \multicolumn{1}{c|}{B-} &
  \multicolumn{1}{c|}{A} &
  \multicolumn{1}{c|}{A+} &
  \multicolumn{1}{c|}{A-} &
    \multicolumn{1}{c|}{Dust Mass ($logM_{\odot}$)} \\
\hline
  Arp220 & 0.018 & 35.51 & 0.25 & 0.26 & 1.58 & 0.01 & 0.01 & 0.93 & 0.01 & 0.01&8.3\\
  NGC6240 & 0.024 & 41.36 & 0.18 & 0.16 & 1.41 & 0.01 & 0.01 & -0.04 & 0.01 & 0.01&7.7\\
  IRAS13120-5453 & 0.031 & 36.78 & 0.07 & 0.07 & 1.64 & 0.01 & 0.00 & 0.48 & 0.01 & 0.01&8.4\\
  Mrk273 & 0.038 & 36.34 & 0.32 & 0.33 & 1.89 & 0.01 & 0.01 & 0.15 & 0.01 & 0.01&8.1\\
  UGC5101 & 0.039 & 34.47 & 0.16 & 0.14 & 1.51 & 0.01 & 0.02 & 0.17 & 0.01 & 0.01&8.4\\
  Mrk231 & 0.042 & 36.87 & 0.41 & 0.43 & 1.89 & 0.02 & 0.02 & 0.26 & 0.02 & 0.02&8.3\\
  IRAS05189-2524 & 0.043 & 46.73 & 0.29 & 0.28 & 1.44 & 0.01 & 0.02 & -0.58 & 0.01 & 0.01&7.7\\
  IRAS20551-4250 & 0.043 & 45.43 & 0.21 & 0.20 & 1.58 & 0.01 & 0.01 & -0.58 & 0.01 & 0.01&7.6\\
  IRAS17208-0014 & 0.043 & 38.12 & 0.20 & 0.18 & 1.66 & 0.01 & 0.01 & 0.28 & 0.01 & 0.01&8.5\\
  IRAS10565+2448 & 0.043 & 36.60 & 0.12 & 0.11 & 1.67 & 0.01 & 0.01 & -0.03 & 0.01 & 0.01&8.2\\
  IRAS23128-5919 & 0.045 & 42.25 & 0.32 & 0.31 & 1.72 & 0.02 & 0.02 & -0.48 & 0.01 & 0.01&7.7\\
  Mrk463 & 0.05 & 43.39 & 3.84 & 4.28 & 1.46 & 0.15 & 0.11 & -1.26 & 0.14 & 0.11&7.2\\
  IRAS08572+3915 & 0.058 & 51.45 & 1.20 & 1.10 & 1.75 & 0.05 & 0.05 & -1.13 & 0.03 & 0.04&7.3\\
  IRAS15250+3609 & 0.058 & 47.67 & 1.06 & 0.97 & 1.45 & 0.03 & 0.04 & -0.90 & 0.03 & 0.03&7.7\\
  IRAS09022-3615 & 0.06 & 35.38 & 0.16 & 0.15 & 2.06 & 0.01 & 0.01 & -0.03 & 0.01 & 0.01&8.3\\
  IRAS19254-7245 & 0.062 & 35.97 & 0.24 & 0.24 & 1.76 & 0.02 & 0.02 & -0.35 & 0.01 & 0.01&8.2\\
  IRAS23365+3604 & 0.064 & 36.49 & 0.19 & 0.20 & 1.85 & 0.02 & 0.02 & -0.26 & 0.01 & 0.01&8.2\\
  IRAS14378-3651 & 0.067 & 37.89 & 0.35 & 0.35 & 1.89 & 0.03 & 0.03 & -0.42 & 0.02 & 0.02&8.1\\
  IRAS22491-1808 & 0.078 & 54.63 & 0.32 & 0.35 & 1.11 & 0.01 & 0.01 & -1.22 & 0.01 & 0.01&7.8\\
  IRAS06035-7102 & 0.079 & 36.91 & 0.69 & 0.71 & 1.94 & 0.05 & 0.06 & -0.46 & 0.04 & 0.04&8.2\\
  IRAS14348-1447 & 0.083 & 37.63 & 0.21 & 0.21 & 1.71 & 0.02 & 0.02 & -0.36 & 0.01 & 0.01&8.5\\
  IRAS19297-0406 & 0.086 & 36.39 & 0.23 & 0.21 & 1.84 & 0.02 & 0.02 & -0.24 & 0.01 & 0.01&8.6\\
  IRAS20414-1651 & 0.087 & 37.34 & 0.58 & 0.66 & 1.65 & 0.03 & 0.03 & -0.52 & 0.03 & 0.02&8.4\\
  IRAS06206-6315 & 0.092 & 36.72 & 0.37 & 0.39 & 1.72 & 0.03 & 0.03 & -0.51 & 0.02 & 0.02&8.4\\
  IRAS08311-2459 & 0.099 & 37.41 & 0.67 & 0.73 & 1.81 & 0.03 & 0.03 & -0.50 & 0.03 & 0.03&8.5\\
  IRAS15462-0450 & 0.1 & 33.60 & 0.99 & 1.07 & 2.28 & 0.09 & 0.09 & -0.63 & 0.07 & 0.06&8.1\\
  IRAS11095-0238 & 0.106 & 41.15 & 1.31 & 1.37 & 2.08 & 0.08 & 0.08 & -1.01 & 0.06 & 0.06&7.9\\
  IRAS20087-0308 & 0.106 & 34.11 & 0.43 & 0.49 & 1.88 & 0.03 & 0.02 & -0.24 & 0.02 & 0.02&8.8\\
  IRAS23230-6926 & 0.107 & 39.11 & 0.77 & 0.74 & 1.99 & 0.05 & 0.05 & -0.79 & 0.04 & 0.04&8.2\\
  IRAS01003-2238 & 0.118 & 45.30 & 2.48 & 2.37 & 1.96 & 0.13 & 0.13 & -1.38 & 0.09 & 0.09&7.7\\
  IRAS12071-0444 & 0.128 & 38.57 & 1.00 & 1.03 & 1.99 & 0.07 & 0.07 & -0.92 & 0.05 & 0.05&8.3\\
  IRAS00188-0856 & 0.128 & 35.76 & 0.88 & 1.05 & 1.74 & 0.05 & 0.05 & -0.67 & 0.05 & 0.04&8.6\\
  IRAS20100-4156 & 0.13 & 39.51 & 0.40 & 0.39 & 1.96 & 0.03 & 0.03 & -0.63 & 0.02 & 0.02&8.6\\
  IRAS23253-5415 & 0.13 & 37.67 & 0.67 & 0.72 & 1.46 & 0.03 & 0.03 & -0.74 & 0.03 & 0.03&8.7\\
  IRAS03158+4227 & 0.134 & 42.30 & 1.26 & 1.37 & 1.58 & 0.04 & 0.04 & -0.84 & 0.04 & 0.04&8.6\\
  IRAS16090-0139 & 0.134 & 37.50 & 0.43 & 0.44 & 1.84 & 0.03 & 0.03 & -0.60 & 0.02 & 0.02&8.7\\
  IRAS10378+1109 & 0.136 & 42.04 & 1.06 & 1.08 & 1.61 & 0.06 & 0.06 & -1.13 & 0.04 & 0.04&8.3\\
  IRAS07598+6508 & 0.148 & 40.34 & 0.83 & 0.81 & 1.60 & 0.05 & 0.05 & -1.09 & 0.04 & 0.04&8.5\\
  IRAS03521+0028 & 0.152 & 42.62 & 1.15 & 1.27 & 1.57 & 0.05 & 0.05 & -1.02 & 0.04 & 0.04&8.6\\
  Mrk1014 & 0.163 & 46.89 & 1.10 & 1.14 & 1.50 & 0.07 & 0.06 & -1.30 & 0.04 & 0.04&8.4\\
  IRAS00397-1312 & 0.262 & 41.73 & 1.19 & 1.19 & 2.03 & 0.08 & 0.08 & -1.21 & 0.05 & 0.05&8.9\\
\hline\end{tabular}
\caption{Results of single component parametric fits to the dust SEDs of HERUS ULIRGs. T+ and T- represent the 68\% confidence interval on the marginalsed MCMC fit to the temperature parameter and similarly for B+, B- and A+ and A- for the $\beta$ and $\log{A}$ parameters respectively. The $\log{A}$ parameter is arbitrarily normalised.
}
\label{table:fits}
\end{table*}

\begin{figure}
\epsfig{file=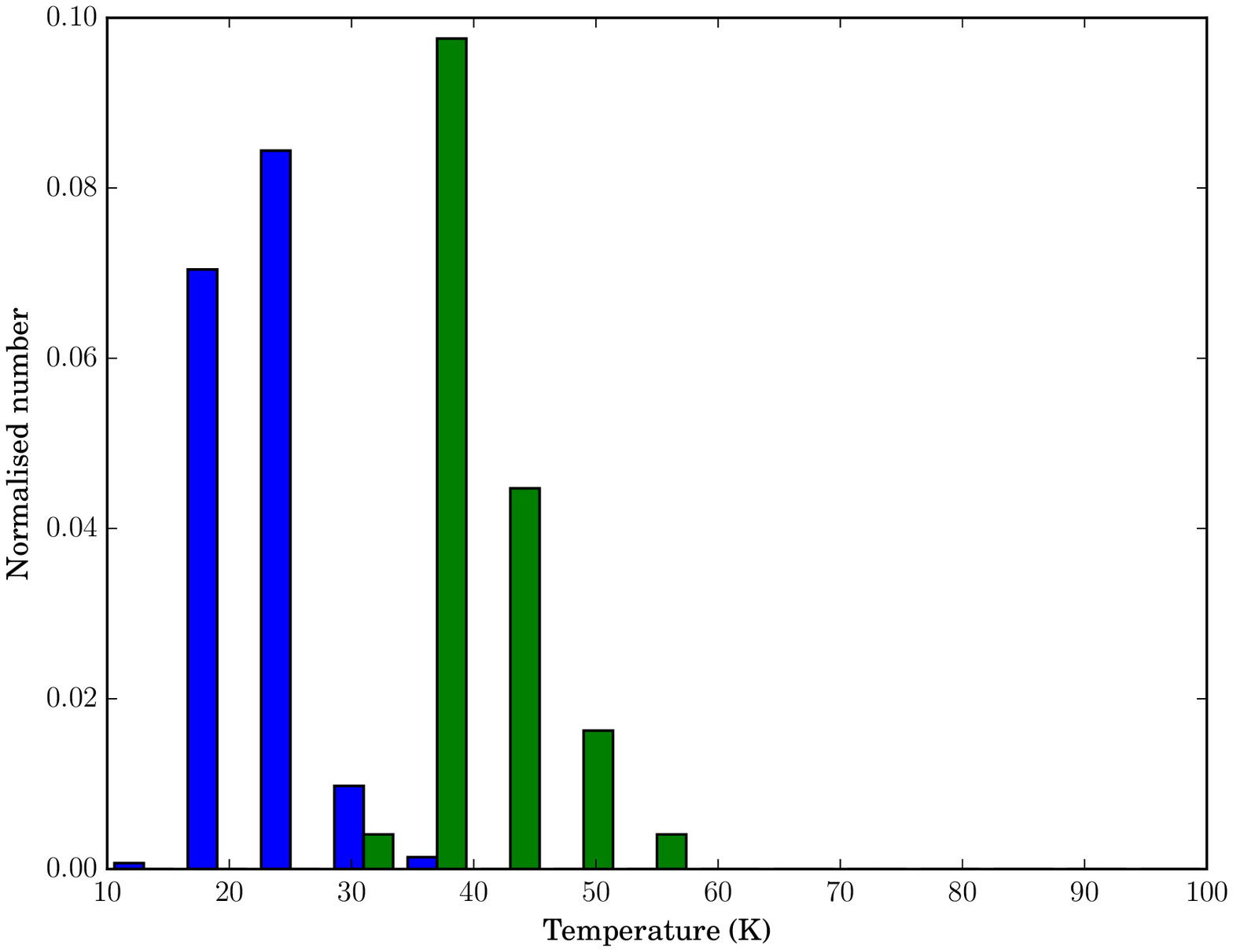,angle=0,width=9cm}
\caption{Histogram of the SED fit temperature parameter from the HERUS ULIRG sources (green) compared to similar results form the lower luminosity HRS galaxies (blue, Cortese et al., 2014).}
\label{fig:temps}
\end{figure}

\begin{figure}
\epsfig{file=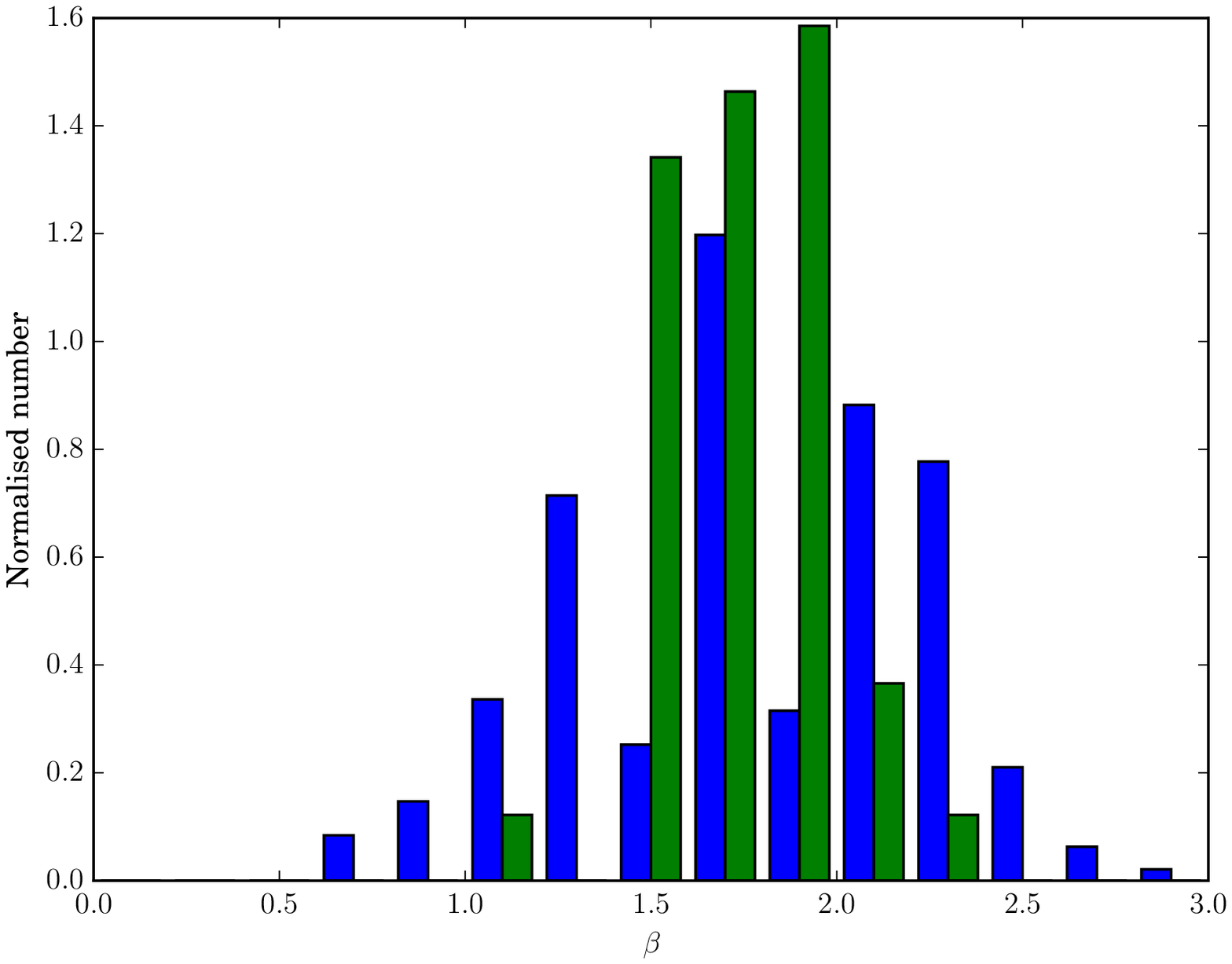,angle=0,width=9cm}
\caption{Histogram of the SED fit $\beta$ parameter from the HERUS ULIRG sources (green) compared to similar results form the lower luminosity HRS galaxies (blue, Cortese et al., 2014).}
\label{fig:beta}
\end{figure}

\subsection{Fit Results}

Plots of the fitted SEDs compared to the observational data can be found in Appendix B. The distribution of temperature and $\beta$ values derived from these fits are shown in Figs. \ref{fig:temps} and \ref{fig:beta}, where they are also compared to similar results from the HRS survey (Cortese et al., 2014). The ULIRGs clearly have warmer temperatures than the lower luminosity HRS sources (median temperatures of 37.9K, standard deviation 4.7K for the ULIRGs compared to 21.3K with a standard deviation of 3.4K for the HRS sources). They also have very similar median $\beta$ values (medians of 1.7 and 1.8 respectively), though the HRS sources have a wider range of $\beta$ values. 

Comparing our results with previous ULIRG SED studies such as the Clements et al. (2010), which are based on IRAS data combined with submm photometry at 850$\mu$m, and in some cases 450$\mu$m, shows broad agreement between the fits, with typical differences of less than 5K between the SPIRE and submm derived temperatures. Deriving population averages from such samples, though, is difficult since previous work is largely based on incomplete samples with non-detections at submm for many of the fainter objects. The SPIRE derived temperatures are usually lower than the submm temperatures. There are three outliers where there is a large difference in derived temperature, with T$_{SPIRE}$ - T$_{submm} > -10$K. One of these (IRAS08572+3915), which has the largest temperature discrepancy in the fitting (T$_{SPIRE}$ - T$_{submm} = -12.5$ K), had only a very low significance 850 $\mu$m detection (2.7$\sigma$) so we might expect its submm SED fit to be unreliable. IRAS 10378+1108, with T$_{SPIRE}$ - T$_{submm} = -11$ K, has a 4$\sigma$ 850$\mu$m detection which, while comparatively low, was not unusual in the Clements et al. (2010) dataset. The final discrepant object, with T$_{SPIRE}$ - T$_{submm} = -10.5$K, is Arp220, which is strongly detected in all bands. Rangwala et al. (2011) used a wider range of photometric data to show that Arp220 is optically thick to  235$\mu$m, which is likely the origin of the discrepancy. 

\subsection{Temperature-$\beta$ Relation}

The relation between dust temperature and $\beta$ is shown in Fig. \ref{fig:tb}, which also shows similar data for the HRS from Cortese et al. (2014). As can be seen the  HRS sources and the HERUS ULIRGs both show an anticorrelation between T and $\beta$, with hotter sources having a shallower $\beta$. The correlation coefficient for the HERUS sources is -0.51 with a probability of this arising at random of 0.00062. However, the relationships are quite different. The gradient of the anticorrelation is significantly shallower for ULIRGs, and there is a clear offset in temperature of about +15K between the HERUS and HRS samples. The temperature offset might simply come from increased dust heating in the higher luminosities of ULIRGs relative to the HRS galaxies, or possibly result from an AGN contribution to the luminosity, but the different relationship to $\beta$ is more interesting. Cortese et al. (2014) find a correlation between $\beta$ and metallicity with lower metallicity systems having a systematically lower $\beta$ value. Temperature, in contrast, was found to be mainly related to specific star formation rate.
Pereira-Santaella et al. (2017) have recently suggested that ULIRGs may be low metallicity systems for their high stellar mass on the basis of far-IR spectroscopy. The range of $\beta$ values seen here for ULIRGs, 1.5 to 2, would, when compared to the HRS sample, imply a fairly typical HRS metallicity of $12 + \log{O/H} \sim 8.5 - 8.7$.  This is consistent with 
the metallicities of $8.5 < 12 + \log{O/H} < 8.9$ inferred by Pereira-Santaella et al. (2017). The higher temperatures seen in ULIRGs for these $\beta$ values would then be due to the higher specific star formation rate in these systems. Alternatively, the anticorrelation may reflect differences in the optical properties of the dust as it evolves. Two studies in our own galaxy (K\"{o}hler et al., 2015; Ysard et al., 2015) show that variations in the dust optical properties due to coagulation are able to reproduce the observed variations in T and $\beta$ from the diffuse to the dense interstellar medium.

\begin{figure}
\epsfig{file=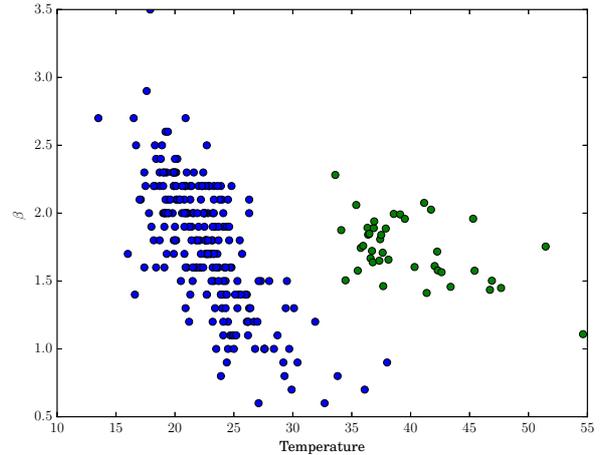,angle=0,width=9cm}
\caption{The temperature-$\beta$ correlation for HERUS ULIRGs (green) and HRS galaxies (blue). }
\label{fig:tb}
\end{figure}

\subsection{Dust Masses}

We follow Cortese et al. (2014) in calculating an estimate of the dust mass in these objects using the equation:
\begin{equation}
F_{\nu} = \frac{M_{Dust}}{D^2} \kappa_{\nu_0} \left( \frac{\nu}{\nu_0}\right)^\beta B_{\nu}(T)
\end{equation}
where $F_{\nu}$, $T$, $\beta$ and $B_{\nu}$ have previously been defined, $D$ is the luminosity distance of the source, $M_{Dust}$ is the dust mass and we take $\nu_0$ to be 856.5 GHz (ie. 350$\mu$m) and $\kappa_{\nu_0}$ to be 0.192 m$^2$ kg$^{-1}$ (Draine 2003). The resulting HERUS ULIRG dust masses, and a comparison to the HRS dust masses from Cortese et al. (2014), are shown in Figure \ref{fig:mass}. The HERUS ULIRGs are found to have significantly larger dust masses than the HRS sources, with a median dust mass of $10^{8.3}$M$_{\odot}$, with standard deviation of $10^{0.4}$M$_{\odot}$, compared to the HRS median of $10^{6.8}$M$_{\odot}$, with standard deviation of $10^{0.5}$M$_{\odot}$. If one considers that ULIRGs are likely the result of a merger of two galaxies and divide their dust mass by 2, giving a median progenitor dust mass of $10^{8.0}$M$_{\odot}$, this is still over a factor of 10 greater dust mass than is seen in the HRS population, and comparable to the most dust rich sources in either the HRS or in the IR-selected samples of nearby galaxies in Dunne et al. (2000) and Vlahakis et al. (2005).
The stellar masses of HRS sources cover a wide range, ($8\leq \log{M_*/M_{\odot}} \leq 12$). Determining the stellar mass of ULIRGs is complicated by the range of obscurations in these systems and by contributions from AGN to the optical/NIR SED. Nevertheless, Hou et al. (2011) used the stellar population fitting code STARLIGHT to derive stellar masses for a sample of 160 ULIRGs selected from SDSS. They found that ULIRG stellar masses range from $10^{10} - 10^{12} M_{\odot}$, with AGN dominated ULIRGs having the highest masses. This places them at the upper range of stellar masses in HRS. The progenitors of ULIRGs must thus be either very massive systems or be among the most dust rich, and thus gas rich, members of the field galaxy population, unless there is significant dust production in the early stages of the evolution of a merger into a ULIRG.

\begin{figure}
\epsfig{file=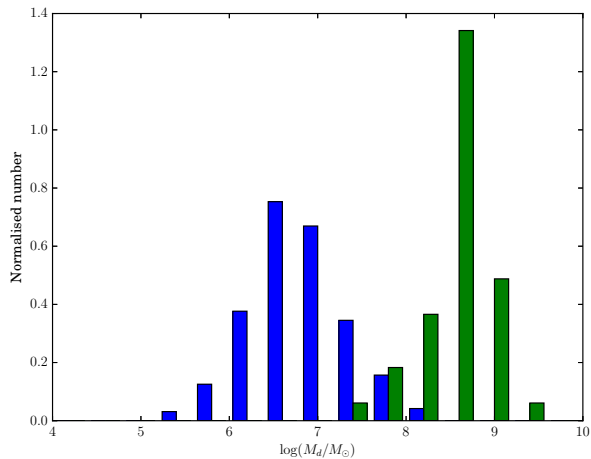, angle=0, width=9cm}
\caption{Histogram of dust masses derived from our SED fits to the HERUS ULIRGs (green) compared to similar results form the lower luminosity HRS galaxies (blue, Cortese et al., 2014).}
\label{fig:mass}
\end{figure}

\subsection{Temperature-Luminosity Diagram}

In Figure \ref{fig:l-t} we plot the positions of the HERUS ULIRGs in the luminosity-temperature plane alongside various comparison samples from the literature. The latter include local galaxies from the HRS (Cortese et al., 2014), SMGs from Chapman et al. (2005), intermediate redshift ULIRGs from Yang et al. (2007) and Magdis et al. (2014), and $z>4$ SMGs from Huang et al. (2014). We also plot the population of H-ATLAS 250$\mu$m selected galaxies from Smith et al. (2013) as contours (sources selected to be detected at $>5\sigma$ at 250$\mu$m, with optical counterparts and spectroscopic redshifts largely from SDSS (Abazajian et al. 2009) or GAMA (Driver et al. 2011) and with $z<0.5$;  see the Smith et  al. paper for details). Most of these works are applying a single temperature T-$\beta$ fitting approach to the data similar to that applied here, so we are comparing like-with-like in this plot.

Previous realisations of this plot, with smaller data sets, poorer coverage of the far-IR SED (eg. Clements et al., 2010), and, in at least some cases, with different assumptions about the dust SED (eg. fixed rather than variable $\beta$ values), suggested that high redshift sources such as those of Chapman et al. (2005), lying at $z\sim$2, or those of Huang et al. (2014), lying at $z>4$, occupied part of this diagram largely disjoint from more local high $L_{FIR}$ sources, with the SMGs having lower temperatures than nearby ULIRGs. The current data sets still show this to some extent, but the gap between the SMGs and more local ULIRGs is being filled, largely by sources from HERUS and Magdis et al. (2014). The highest redshift SMGs in this plot, from Huang et al. (2014), also appear rather warmer than the lower redshift SMGs of Chapman et al.. This suggests that the earlier distinction between the two populations was likely a result of sample incompleteness, with earlier observations failing to cover cooler local ULIRGs or warmer high $z$ SMGs. There is also the possibility that the different selection methods used for these sources - 60$\mu$m flux for {\em IRAS}-based samples like HERUS, 250$\mu$m flux for the {\em Herschel}-based samples such as Magdis et al., and 850$\mu$m flux for the SMGs of Chapman et al. and Huang et al., might have an effect. This is a possibility we explore in more detail later in this paper.

\begin{figure*}
\epsfig{file=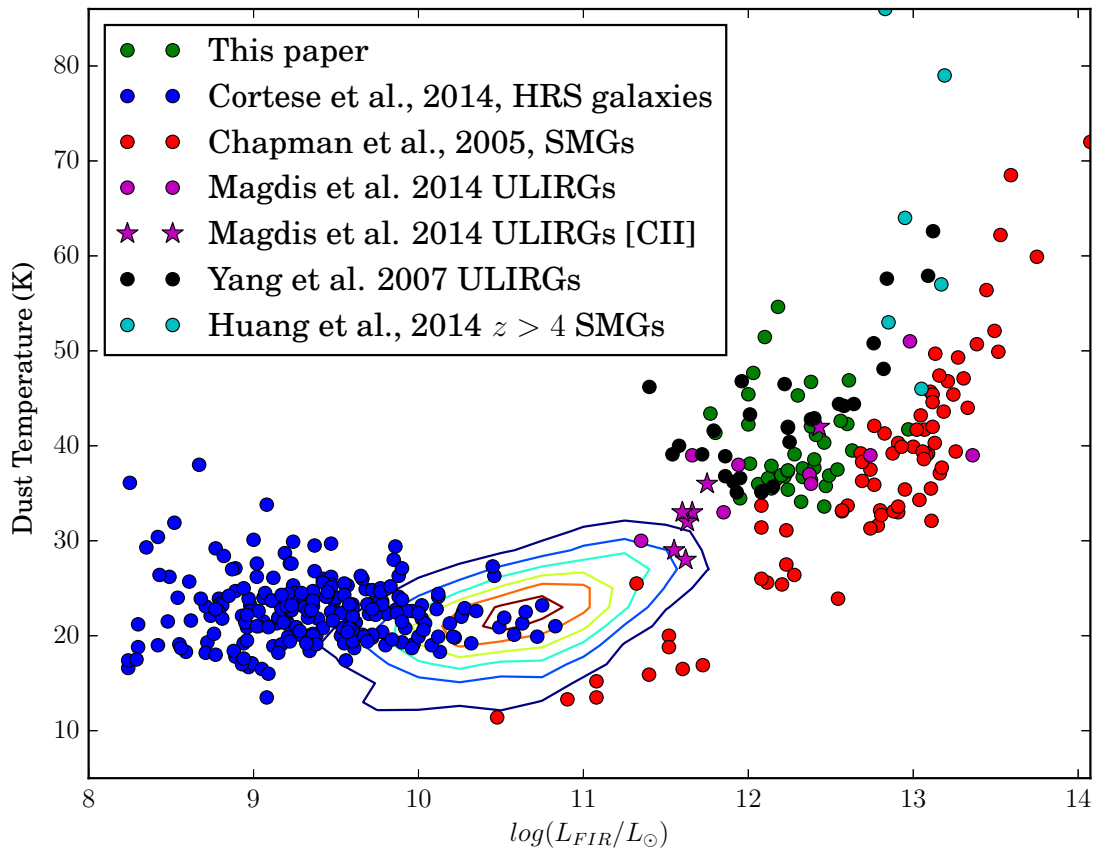, angle=0, width=17cm}
\caption{The FIR luminosity-temperature plane for the HERUS ULIRGs and other samples from the literature. HERUS ULIRGs are shown as green dots while sources from he HRS (Cortese et al., 2014) are shown in blue, SMGs from Chapman et al. (2005) are shown in red, intermediate redshift ULIRGs from Yang et al. (2007) are shown in black and Magdis et al. (2014) shown in magenta (those sources with [CII] observations are shown as stars, the others as dots), while $z>4$ SMGs from Huang et al. (2014) are shown in cyan. The contours indicate the distribution of 13826 H-ATLAS 250$\mu$m selected sources from Smith et al. (2013).}
\label{fig:l-t}
\end{figure*}

\section{Discussion}

\subsection{What are the real Dust SEDs of ULIRGs?}

While we have fitted a parameterised single $(T,\beta)$ model to the SEDs of the HERUS galaxies, it is far from certain that this simple model is adequate for fitting the data. A single temperature and $\beta$ value for all of the dust in a galaxy is almost certainly an over simplification, especially since physical dust properties are likely to evolve differently in different environments in the same galaxy (see eg. K\"{o}hler et al., 2015). Our current investigation uses only the five photometric points from {\em Herschel} and {\em IRAS}. More complex models can be fitted to those sources where more data is available - an example for this can be found for Arp220 in Rangwala et al. (2011). Such an analysis will be the subject of a future HERUS paper, but this additional photometric data is inhomogeneously distributed, with some HERUS sources being very well studied while others having little more data than is presented here. Conclusions from such a study might thus be biased in some statistical sense towards, for example, brighter or more nearby ULIRGs. The use of a simple model that can be applied uniformly to a larger data set, as done here, can thus have substantial value. 

We can gain insights into the appropriateness of our simple parametric SED model by searching for trends in the flux residuals between our fits and the real data. We plot the relative residuals, $(F_{model} - F_{obs})/F_{obs}$ against model dust temperature in Figure \ref{fig:residuals}.

The SPIRE and {\em IRAS} data behave quite differently. The model generally underpredicts SPIRE fluxes by about 20\%, but there is no correlation between the residual and the modelled dust temperature - the results of fitting a straight line to the data finds correlation coefficients of -0.06, 0.13 and 0.07 with the probability of there being no correlation of 0.72, 0.43 and 0.65 for the 250, 350 and 500$\mu$m channels respectively. In contrast, the model usually overpredicts the 60 and 100$\mu$m fluxes, and the size of this residual, ranging from 0 to about 25\%, is correlated with model dust temperature. This effect is most apparent at 100$\mu$m and can be seen in many of the SED fits plotted in Appendix B. The correlation coefficients  are -0.3 and -0.6, and the chances of this arising at random are 0.05 and $2 \times 10^{-5}$ at 60 and 100$\mu$m respectively. A correlation between temperature and residual is thus clearly present at 100$\mu$m and marginally so at 60$\mu$m, but completely absent in the {\em Herschel} bands.

This result suggests that the real SEDs of ULIRGs may be different to the simple parametric model used here. This is not a surprise in itself since there are several ways in which the SED of a genuine object is likely to differ from a single temperature parameterization. These include the presence of dust components at other temperatures, whether cooler than the temperature found in our single $T$ fits (eg. Clements et al., 2010; Dunne \& Eales, 2001), the contamination of the 60$\mu$m band by significantly warmer dust (eg. Mrk1231 and Mrk273), or a flattening of the 60 to 250$\mu$m SED as a result of the dust being optically think at these wavelengths, as has been found for Arp220 by Rangwala et al. (2011). A full analysis of these possibilities would require more data points than the five we have for the complete HERUS sample. However we
we may gain some insight into these more complex models by seeing if the amplitude of the residuals and the correlation with temperature can be reproduced by simple toy models of more complicated dust SEDs.

We explore three different possibilities, (i) hot dust contamination at short wavelengths, (ii) cold dust contribution rising to longer wavelengths, and (iii) dust that is optically thick at far-IR/submm wavelengths, by producing model HERUS-like catalogs for the various possibilities and then fitting a single temperature $(T,\beta)$ model to them using the same MCMC code described above. The simulated HERUS fluxes have gaussian noise added to them matching the average S/N ratio of the appropriate HERUS band before the fitting process.

The possibility of hot dust contamination at short wavelengths is modelled by adding a 100K $\beta=1.7$ dust component to an underlying $(T,\beta=1.7)$ SED with temperatures ranging from 30 to 60K. $\beta=1.7$ is chosen since this is the average $\beta$ value for the HERUS sample. The strength of this hot dust contribution was varied up to a maximum of 50\% of the flux in the 60$\mu$m band. Cold dust contamination is modelled in a similar way, by adding a 20K $\beta=1.7$ component to a similar range of underlying templates, but with the contribution normalised to some fraction of the 250$\mu$m flux, rising once again to a maximum of 50\% of the 250$\mu$m flux. The possibility of optically thick dust was modelled with a thermal dust SED with temperature ranging from 30 to 60K and a $\beta$ of 1.7 but which has an optical depth of 1 at 200$\mu$m. This is similar to the dust SED found for Arp220 by Rangwala et al. (2011).

None of these simple, uniform, models for the underlying SED reproduces what is seen in the data. Neither of the hot or cold additional dust component models can reproduce fractional residuals as large as those seen, even when the additional component makes up 50\% of the flux of the appropriate normalisation band (60$\mu$m for hot dust and 250$\mu$m for cold). The optically thick model also fails to reproduce the size of the residuals. Furthermore, none of the models could reproduce the pattern of residuals, where we see similar, negative, residuals in the three SPIRE bands that are all uncorrelated with fitted temperature, and positive residuals correlated with temperature in the IRAS bands. Better sampling of the far-IR SEDs of ULIRGs so that we can determine the  properties of each individual object will be necessary to make progress on this problem.

\begin{figure}
\epsfig{file=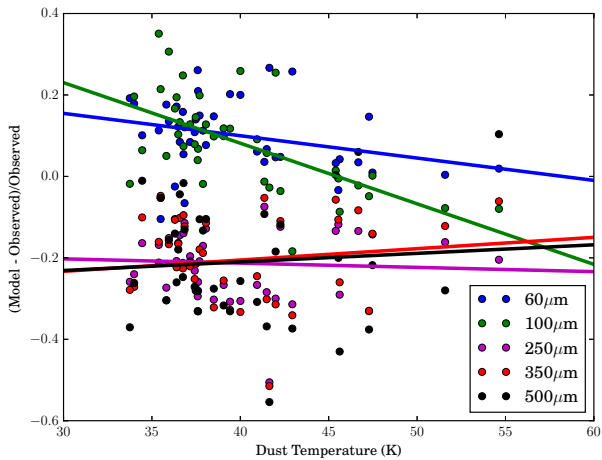, angle=0, width=9cm}
\caption{Flux residuals, (model flux - observed flux)/observed flux, for the 60, 100, 250, 350 and 500$\mu$ bands (blue, green, magenta, red, black respectively) plotted against fitted dust temperature. The solid lines are straight line fits to the data at the matching wavelength. Correlations between residual flux and fitted temperature are only found for the 60 and 100$\mu$m channels, of which the 100$\mu$m correlation is by far the strongest.}
\label{fig:residuals}
\end{figure}

\subsection{Far-IR Colours}

While fitting models to the SEDs of {\em Herschel} sources is one way to compare different populations, another is simply to compare their far-IR colours. In Figure \ref{fig:colours} we show the 100-to-250 $\mu$m and 350-to-500 $\mu$m colours of the HERUS sources, HRS sources, and those Magdis et al. (2014) sources where 100$\mu$m fluxes are available. The H-ATLAS sources lie in the same region as the bulk of the HRS Spiral-type galaxies. However, the HERUS ULIRGs lie in an almost completely separate part of the colour-colour diagram to the HRS sources or the Magdis et al. ULIRGs. 
Only one HRS source lies in the HERUS ULIRG region -  NGC4697, classified as a post-merger remnant elliptical (eg. Arnold et al., 2014) hosting a low luminosity AGN (Yuan et al., 2009). Similarly, only one Magdis ULIRG for which 100$\mu$m fluxes are available, CDFS2, lies close to the HERUS ULIRG region. The Magdis et al. ULIRGs with measured 100$\mu$m fluxes plotted here lie at redshifts from 0.21 to 0.35, while the highest redshift HERUS ULIRG is at z=0.26. The two samples thus overlap in redshifts, so K-corrections and evolution will not be responsible for the differences we are seeing. The HERUS ULIRGs were selected at 60$\mu$m, while the Magdis ULIRGs were selected at 250$\mu$m. We also note that the HRS spirals have comparable colours to the overall population of 250$\mu$m selected galaxies in H-ATLAS.

We also plot colour tracks for two simple $(T,\beta)$ models, one with $\beta=2$ and the other with $\beta=1$ and temperature ranging from 15-45K for the former and 15-75K for the latter. Most galaxies in this plot appear to be scattered between these two lines with the HRS and Magdis sources having lower temperatures than the HERUS ULIRGs. A small fraction of HRS sources, largely E and S0 types, lie below the $\beta$=1 line, as does one of the Magdis ULIRGs and two of the HERUS sources. These latter are 3c273 and IRAS13451+1232 whose far-IR SEDs are dominated by non-thermal emission. This may well be the case for the other sources in this part of the diagram. The most extreme HERUS sources on the right of this diagram are IRAS08572+3915 and IRAS 01003-2238, which both lie on a rightwards extension of the $\beta=1$ track to higher temperatures (though their best SED fits have $\beta$ values closer to 2). Both of these sources are known to contain dust obscured AGN (Imanishi et al. 2007; Efstathiou et al., 2014). The other HERUS source lying on the $\beta=1$ line at somewhat lower temperature is Mrk463 which also hosts an AGN.

It is tempting to envisage a potential evolutionary track in this diagram whereby two cooler dust galaxies close to the locus of HRS spiral-type galaxies merge, triggering a starburst which warms their dust and moves the merging system towards the HERUS ULIRG region along the $\beta=2$ line. The merger also funnels gas into the nuclear regions of the merging galaxy, fuelling an AGN. As the starburst ages and the AGN becomes more significant the merger moves away from the $\beta=2$ line towards the $\beta=1$ line. In this context it is interesting to note that the Magdis et al. ULIRGs have a similar $L_{[CII]}/L_{FIR}$ ratio to normal star forming galaxies, which they are closer to in this colour-colour diagram, than do local IRAS selected ULIRGs, which appear to be [CII] deficient (Farah et al., 2013).

\begin{figure*}
\epsfig{file=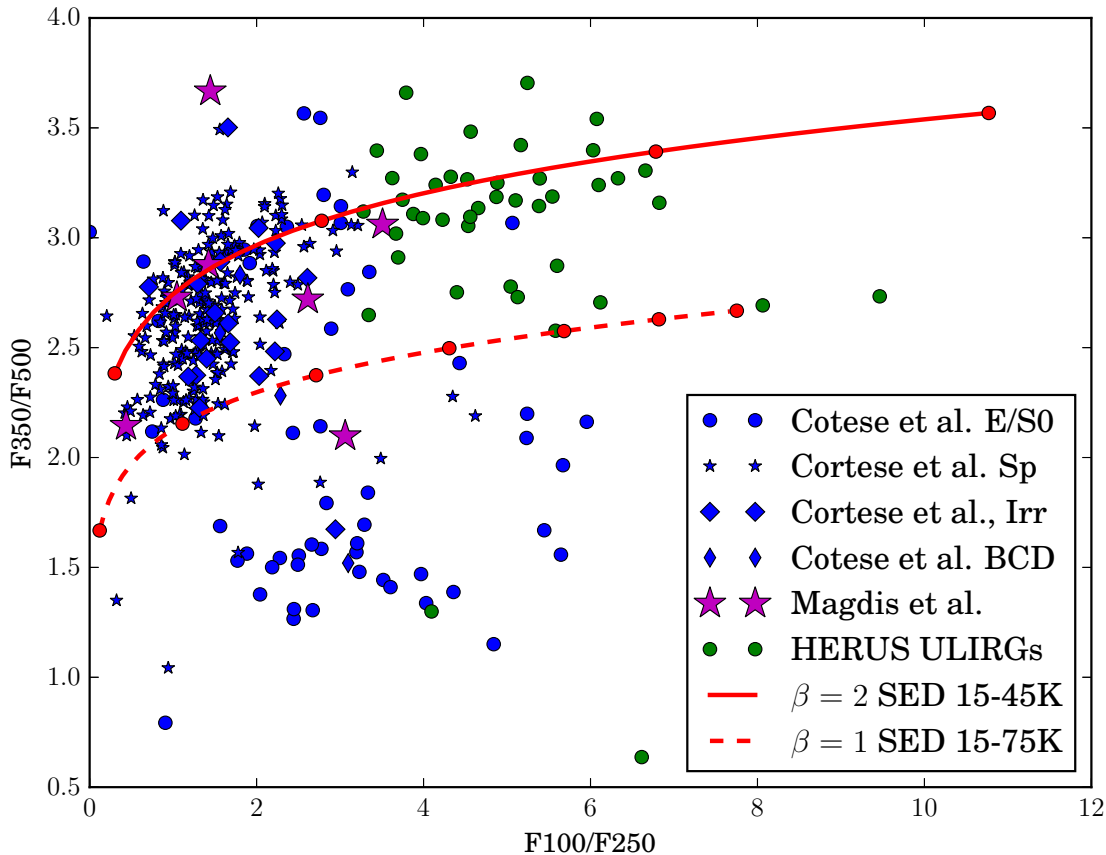,angle=0, width=17cm}
\caption{The colours of HERUS ULIRGs compared to other samples. F100/F250 vs F350/F500 shown for HERUS ULIRGs (green dots), HRS galaxies (blue symbols) of different morphological types (E \& S0 dots, Spiral stars, Irr and Pec upside diamonds, blue compact dwarfs narrow diamonds), and Magdis ULRGs where data is available (magenta stars). Also plotted are lines showing the colours of modified black body spectra with $\beta$ = 2 (solid) from 15 to 45K and $\beta$ = 1 (dashed) going from 15K to 75K. Solid dots along these lines indicate temperatures in increments of 10K starting at 15K at the left of the plot. The H-ATLAS 250$\mu$m sources occupy the same region of the diagram as the bulk of the Cortese et al. (2014) HRS sources. Note that the IRAS-selected ULIRGs and the Magdis 250$\mu$m selected sources have an almost disjoint range of colours, and that the Magdis sources have colours much more typical of local star forming galaxies from HRS.}
\label{fig:colours}
\end{figure*}

\subsection{ULIRG selection functions}

While evolutionary state and/or AGN content is a possible explanation for the difference between the HERUS and Magdis ULIRGs, another plausible explanation is the possibility that the different wavelengths at which these sources are selected - 60$\mu$m in the case of the HERUS ULIRGs and 250$\mu$m in the case of the Magdis ULIRGs, and the H-ATLAS galaxies which have similar colours - might lead to systematic biases in the types of sources found by these different surveys. To examine this we select two of the panchromatic SEDs for {\em Herschel} sources provided by Berta et al. (2013), one matching the average colour of the HERUS ULIRGs and the other matching that of the HRS sources and Magdis ULIRGs. We then look at how the observed flux of such a template source with a luminosity of $10^{12} L_{\odot}$ varies with redshift and compare this to the selection criteria used for HERUS at 60$\mu$m and in Magdis et al. (2013) at 250$\mu$m.

The Berta templates selected were the cold galaxy (template 2 in the Berta template library) for 250$\mu$m selected sources, matching average colours of F100/F250 $\sim$1.5 and F350/F500 $\sim$ 2.7, and the obscured star forming galaxy (template 8 in the template library) for IRAS selected ULIRGs matching their average colour of F100/F250 $\sim$ 5 and F350/F500 $\sim$ 3. There is substantial scatter in the colours of these different samples so these specific templates will not give a perfect match to the data. Nevertheless, they can provide insights into any major differences resulting from the selection of sources with such different colours at different wavelengths (see also Fig. 5 of Magdis et al., 2013).

\begin{figure*}
\epsfig{file=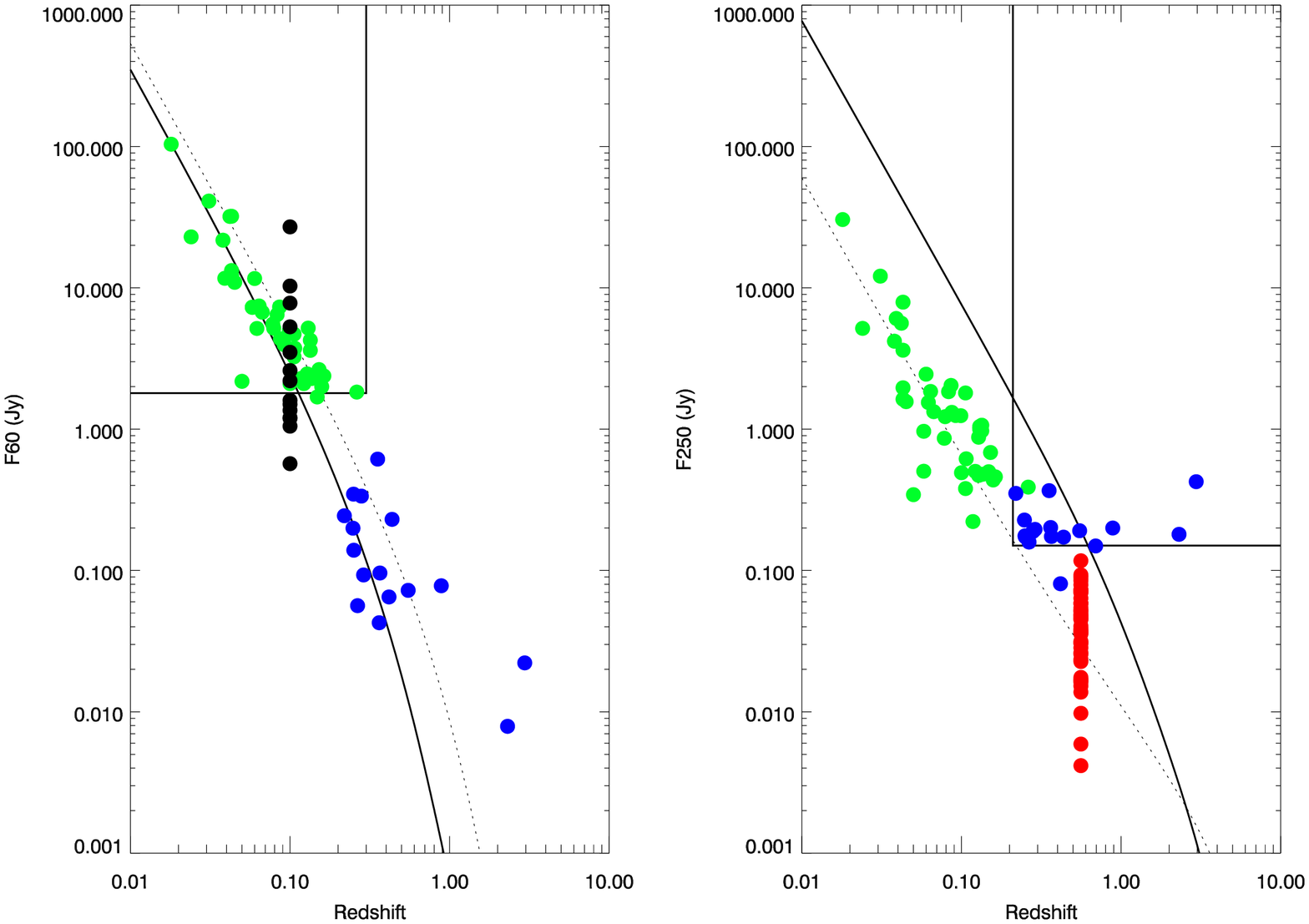,angle=90, width=15cm}
\caption{Flux of a 10$^{12} L_{\odot}$ galaxy as a function of redshift at 60$\mu$m (left) and 250$\mu$m (right) as derived from the Berta et al. (2013) templates. Solid line is for a cold galaxy, appropriate for the 250$\mu$m selected Magdis et al. objects, dotted line is for an obscured star forming galaxy, appropriate for an IRAS selected ULIRG. Dots show the fluxes and redshifts for the HERUS ULIRGs (green) and the Magdis et al. ULIRGs (blue - we use the 70$\mu$m fluxes for these sources since 60$\mu$m fluxes are not available). Solid lines indicate the selection functions of this work on the 60$\mu$m plot and of Magdis et al. on the 250$\mu$m plot. Note that the cold galaxy template is about an order of magnitude brighter at 250$\mu$m than the obscured star forming galaxy at any given redshift for the same luminosity, while at 60$\mu$m the fluxes are about the same. This will lead to the selection of significantly different objects at 250$\mu$m at any given flux limit. The red points in the right hand plot indicate the 250$\mu$m fluxes that the HERUS sources would have if they were redshifted to $z=0.56$. None would be bright enough at $250\mu$m to be selected by Magdis et al. The black points in the left hand plot indicate the 60$\mu$m fluxes of Magdis et al. sources if they were at a redshift of 0.1. Half of the sample, 7 galaxies, would be selected by the HERUS criteria.}
\label{fig:sel1}
\end{figure*}

The flux vs. redshift for these templates for a $10^{12} L_{\odot}$ source, together with the selection functions for the HERUS ($S_{60}>$1.8Jy and $z<0.3$) and Magdis samples ($S_{250}>$0.15Jy and $0.9>z>0.21$), as well as the fluxes and redshifts for the sources in these samples, are shown in Figure \ref{fig:sel1}. As can be seen, for selection at 60$\mu$m the warmer SED sources are marginally favoured since, at any given redshift, they are somewhat brighter than the cold SED sources, by a factor of about 1.5. However, for selection at 250$\mu$m the situation is very different. A cold SED source will be substantially brighter at 250$\mu$m than a hot SED source, by a factor of up to an order of magnitude, out to redshifts of 1-2. A hot SED source would have to be much more luminous to be selected at a given 250$\mu$m flux than a cold SED source. 

Figure \ref{fig:sel1} also shows how selection at 250$\mu$m would affect the HERUS sample if it were shifted to higher redshifts. To do this we use the fitted HERUS SEDs to predict the 250$\mu$m flux of each of our sources if it were shifted to a redshift of 0.56, at the centre of the redshift range of the main Magdis et al. selection. These sources are shown in red. Not one of them would have been bright enough for inclusion in the Magdis sample. Significantly more-luminous IRAS-type, 60$\mu$m selected ULIRGs could still be selected at 250$\mu$m, but since the luminosity function at 250$\mu$m declines steeply at these luminosities (eg. Marchetti et al., 2016) the number density of such sources is very low. We also perform a similar analysis on the Magdis ULIRGs to see how many would be selected by IRAS at 60$\mu$ if they were at the mean redshift of the HERUS sample, $z=0.1$. Here we find that roughly half of the Magdis sources would be detected by IRAS at this redshift.

The co-moving volume of the IRAS survey is much larger than that of the Magdis et al. survey. The parent catalog on which HERUS is based covers an area of 16300 square degrees and our selection reaches out to z=0.3, yielding a total co-moving volume of 3 Gpc$^3$. The Magdis survey covers 100 square degrees and a redshift range of 0.21 to 0.9, for a co-moving volume of 0.3 Gpc$^3$. If we ignore evolution, and assume that 250$\mu$m selected ULIRGs have the same co-moving density in the local universe as they are observed to have in the Magdis survey, there should be 70 such objects seen by IRAS that would have been included in the HERUS survey, when there are in fact none.

However, it has long been known that the ULIRG population strongly evolves with redshift. If we assume that the Magdis ULIRGs evolve at the same rate as the general ULIRG population, whose number density $n$ has been found to evolve as $n \propto (1+z)^n$ with $n = 7.05^{+0.25}_{-0.18}$ (Patel, 2012, which presents an improved determination of this parameter from Patel et al. (2013);  see also Kim \& Sanders, 1998) then their number density is reduced by a factor of 11.75 from $z=0.56$, the mean redshift of the Magdis survey, to 0.1, the mean redshift of the HERUS sample. The expected number of Magdis ULIRGs in the HERUS sample would then be 6. If the two classes were entirely distinct this would represent a $\sim 2.4\sigma$ discrepancy which, while intriguing, is not a formally significant result. However, since there is not a solid boundary between the two classes, with some cross-over at F100/F250 $\sim$ 4 (see Figure \ref{fig:colours}) then the significance of this difference falls still further and we conclude that the Magdis-type ULIRGs are too rare in the local universe for the existing 60$\mu$m selected IRAS surveys to pick them up in significant numbers.

We thus conclude it is possible that the apparent dichotomy drawn between low z {\em IRAS} selected ULIRGs and the higher z, 250$\mu$m selected ULIRGs in Magdis et al. (2014), with the lower redshift objects showing [CII] deficiency, more intense excitation (Rigopoulou et al., 2014) and smaller star forming regions (Diaz-Santos et al., 2013), might be a result of selection at different wavelengths and not indicate very rapid evolution in the properties of the ULIRG population. Searches for ULIRG-luminosity objects with cool dust SEDs in the local universe, possibly using the IRAS 100$\mu$m catalog or the AKARI 140$\mu$m catalog as a starting point, and for warm dust ULIRGs at higher redshifts, possibly using PACS data, may be able to test this idea. Further study of intermediate redshift ULIRG samples (eg. Combes et al., 2011) may also be very useful.

\section{Conclusions}

We present a far-IR photometric atlas of local IRAS selected ULIRGs from the HERUS survey, including fluxes and images at 250, 350 and 500$\mu$m, derived from observations with the SPIRE instrument on the {\em Herschel} Space Observatory together with images of these sources at the three respective wavebands. We fit the SEDs of these sources using a simple optically thin dust SED and derive temperatures and emissivity parameter $\beta$. We find that the dust temperatures in our sources are significantly higher than the dust temperatures of lower luminosity sources, but that the distribution of emissivity parameters are similar. We compare the temperature and luminosity of our sources to those of far-IR emitting galaxies at a wide range of luminosities and redshifts, finding that while high redshift SMGs still appear colder than local ULIRGs of a similar luminosity, the distinction between these samples is becoming less distinct as more sources are observed. A comparison of the far-IR colours of our sources and lower luminosity objects from the literature uncovers an unexpected dichotomy between our ULIRGs, lower luminosity systems, and ULIRGs selected at longer wavelengths using the SPIRE 250$\mu$m band. This difference may result from selection at different wavelengths, with 250$\mu$m selection biased in favour of selecting sources with a cooler SED over a wide range of redshifts.

The next steps for SED studies of local ULIRGs are to use a wider range of existing photometric data to study the far-IR SEDs in more detail through fitting more complex models. Extant other far-IR to submm data, however, unlike that presented in this paper, is inhomogeneous, with some famous sources, such as Arp220 or Mrk231, observed in many more bands than some of the less well-studied objects.



{\bf Acknowledgements}

SPIRE has been developed by a consortium of institutes led
by Cardiff Univ. (UK) and including: Univ. Lethbridge (Canada);
NAOC (China); CEA, LAM (France); IFSI, Univ. Padua (Italy);
IAC (Spain); Stockholm Observatory (Sweden); Imperial College
London, RAL, UCL-MSSL, UKATC, Univ. Sussex (UK); and Caltech,
JPL, NHSC, Univ. Colorado (USA). This development has been
supported by national funding agencies: CSA (Canada); NAOC
(China); CEA, CNES, CNRS (France); ASI (Italy); MCINN (Spain);
SNSB (Sweden); STFC, UKSA (UK); and NASA (USA). HIPE is a joint development by the {\em Herschel} Science Ground Segment Consortium, consisting of ESA, the NASA {\em Herschel} Science Centre, and the HIFI, PACS and SPIRE consortia. This research has made use of the NASA/IPAC Extragalactic Database (NED) which is operated by the Jet Propulsion Laboratory, California Institute of Technology, under contract with the National Aeronautics and Space Administration. E.GA is a Research Associate at the Harvard-Smithsonian
Center for Astrophysics, and thanks the Spanish 
Ministerio de Econom\'{\i}a y Competitividad for support under projects
FIS2012-39162-C06-01 and  ESP2015-65597-C4-1-R, and NASA grant ADAP
NNX15AE56G. JBS wishes to acknowledge the support of a Career Integration Grant within the 7th European Community Framework Program, FP7-PEOPLE-2013-CIG-630861-FEASTFUL. J.A. acknowledges financial support from the Science and Technology Foundation (FCT, Portugal) through research grants PTDC/FIS-AST/2194/2012 and UID/FIS/04434/2013.
DLC and JG acknowledge support from STFC, in part through grant numbers ST/N000838/1 and ST/K001051/1. The authors wish to thank A. Jaffe and D. Smith for useful conversations, and the anonymous referee for their comments.
\\~\\

\begin{appendix}

\section{SPIRE Images of HERUS Sources}

This appendix presents the SPIRE images of the HERUS ULIRGs at 500, 350 and 250$\mu$m. Images will be available in the published paper.

Notes on specific images:

{\bf IRAS03521+0028}: There is significant extended structure in the far-IR sky near this source which can be seen in the background of the SPIRE maps of this object. This is visible in the wider field IRAS 60 and 100$\mu$m maps. Nevertheless, the source is compact and well localised so this structure, likely a result of galactic foregrounds, has no impact on the photometric accuracy for this source.

{\bf IRAS06035-7102}: This image is based on a cut-out of the larger SPIRE survey of the LMC. It thus has a somewhat different format to the other images which show the full SPIRE small scan map around the target ULIRG.

\end{appendix}

\end{document}